\documentclass[%
 reprint,
 amsmath,amssymb,
 prl,
]{revtex4-2}
\usepackage{graphicx}
\usepackage{booktabs}
\usepackage{dcolumn}
\usepackage{bm}
\usepackage{amsmath}

\usepackage{tensor}
\usepackage{geometry} 
\usepackage{circledsteps}
\usepackage[hidelinks]{hyperref}
\usepackage{cleveref}
\usepackage{natbib}
\usepackage{aas_macros}
\begin{document}

\preprint{APS/123-QED}

\title{$CPT$-Symmetric Kähler–Dirac Fermions}

\author{Latham Boyle${}^{1,2}$}
\email{latham.boyle@ed.ac.uk}
\author{Wei-Ning Deng${}^{3,4}$}
\email{wnd22@cam.ac.uk}

\affiliation{
${}^1$Higgs Centre for Theoretical Physics, University of Edinburgh, Edinburgh, EH9 3FD, UK\\
${}^2$Perimeter Institute for Theoretical Physics, Waterloo, Ontario, N2L 2Y5, Canada\\
${}^3$Cavendish Laboratory, J.J. Thomson Avenue, Cambridge, CB3 0HE, UK\\
${}^4$Kavli Institute for Cosmology, Madingley Road, Cambridge, CB3 0HA, UK}



\date{\today}

\begin{abstract}
Kähler–Dirac (KD) spinors have generated excitement in the lattice gauge theory community, as a way to (i) deal with the ``fermion doubling" problems that plague ordinary (Dirac, Majorana, or Weyl) spinors when discretized on a lattice, and (ii) help explain the structure of the standard model.  But if one naively quantizes this theory in Lorentzian signature, problems arise: half the KD fields have the ``wrong sign" Lagrangian, and give rise to negative norm states.  Here we propose a new resolution/interpretation: the KD field actually lives on a two-sheeted spacetime, with the sheets related by $PT$ symmetry or, alternatively,  by $i\leftrightarrow-i$.  And, to avoid any unphysical interactions between the two sheets, the KD field obeys a reality condition (which we call the ``KD-Majorana condition"), which forces every particle on one sheet to be accompanied by a mirror (anti-)particle on the other sheet. We discuss how the standard model fits in this framework, how the fermion (kinetic and Yukawa) terms simplify, and how it may relate to the CPT-symmetric universe model.
\end{abstract}

\maketitle

\textbf{Introduction}: The standard model is a chiral gauge theory (where left- and right-handed fermions transform differently under gauge transformations) \cite{Cheng:1984vwu, Peskin:1995ev, Schwartz:2014sze}.  We know how to treat such theories in perturbation theory, but not how to define them non-perturbatively (and on the lattice in particular) \cite{Montvay:1994cy, Smit_2002,Gattringer:2010zz}. This is an important problem of principle: it suggests that we are missing something important about the geometric meaning of spinors and their relationship to spacetime.  

Wilson's formulation of lattice gauge theory \cite{Wilson:1974sk} works beautifully for bosonic fields, since each such field's geometric meaning dictates how it should live on the lattice: $0$-form (scalar fields) live on $0$-cells (vertices), $1$-forms (gauge fields) lives on $1$-cells (links), $2$-forms (fields strengths) lives on $2$-cells (plaquets), etc; and we can apply the well-developed formalism for discrete exterior calculus on a discrete cell complex (see {\it e.g.}\ \cite{desbrun2005discrete, desbrun2006discrete}).  

However, when spinors enter the story, problems arise: since they do not have a natural $p$-form interpretation, they are traditionally discretized by placing them on the vertices of the lattice/complex; but then, in the continuum limit, one finds more fermions than expected \cite{Montvay:1994cy,Smit_2002}; and, if one starts with a {\it chiral} theory then, after discretizing and taking the continuum limit, one ends up with a {\it non-chiral} theory \cite{Nielsen:1981hk, Nielsen:1980rz, Nielsen:1981xu}.  These are the ``fermion doubling" problems.

Kähler–Dirac (KD) fermions \cite{Kähler1962} are a leading approach to deal with these problems. These are variants of Dirac fermions that also have a dual interpretation in terms of $p$-forms, so that they may be naturally discretized on a lattice, with no discrepancy between the discrete and continuum theories -- both theories share the same (co)homological properties, without unwanted doublers \cite{RABIN1982315, Becher1982TheDE, Montvay:1994cy}.  For a cubic lattice in flat space, this KD approach was shown \cite{Becher1982TheDE} to be equivalent to another popular method called staggered fermions \cite{Kogut:1974ag}; but it is conceptually superior as it generalizes to an arbitrary cell complex in an arbitrary curved spacetime.  Moreover, when one considers chiral or ``restricted" KD fermions, one is naturally led by anomaly considerations to multiplets of four restricted Kähler–Dirac fields which automatically transform like a single generation of fermions in the Pati-Salam unification scheme \cite{Catterall:2022jky, Catterall:2023nww}, raising hope that this improved geometrical picture of fermions may also help explain the symmetry structure of the standard model.

Since simulations are usually done in Euclidean space, KD fermions are usually studied in Euclidean signature. However, given the above motivations suggesting that they are of particular interest, it is natural to consider their quantization and physical properties in Lorentzian signature, which is the subject of this letter.

As we shall review, in Lorentzian space, KD theory naively seems to have a problematic Lagrangian, which has the ``wrong sign" for half of the fields. Here we propose a solution: the fields live on ``two sheets of spacetime" which are related by a $PT$ symmetry or, equivalently, by swapping $i\leftrightarrow-i$.  Then, on both sheets, the states have positive energy and positive norm.  We then discuss how this framework fits the standard model, how the Yukawa terms neatly simplify, and how it relates to the CPT-symmetric universe idea \cite{Boyle:2018tzc, Boyle:2018rgh, Turok:2022fgq, Boyle:2022lcq}.

\vspace{0.25cm}
\textbf{Basics}: In this letter, we adopt the $(+,-,-,-)$ metric signature. The KD action is 
\begin{equation}\label{eq:KD_action}
  S_{KD}=\langle \bar{\Phi}|(K-m)\Phi\rangle=
  \int (\star\bar{\Phi})\wedge(K-m)\Phi
\end{equation}
where $\Phi=\sum_{p=0}^{4}\varphi_{\mu_1\ldots\mu_p}dx^{\mu_1}\wedge\ldots\wedge dx^{\mu_p}$ is a polyform field (a sum of a 0-form, 1-form, 2-form, etc), $\ast$ is the Hodge star operator, and $K=d-d^{\dagger}$, where $d$ is the exterior derivative and $d^{\dagger}=\star d \star$ is its adjoint.  Varying yields the KD equation
\begin{equation}\label{eq:KD_eq}
    (d - d^\dagger - m)\Phi = 0.
\end{equation}
Note that the KD operator $K$ squares to the Hodge-de Rham operator, $-K^2 = dd^\dagger + d^\dagger d = \square$, the wave operator on forms.

In 4D, $\Phi$ is a sum of p-forms with $p=0,\dots,4$, with $1+4+6+4+1=16$ complex components; and the replacement $dx^{\mu}\to\gamma^{\mu}$, assembles these components into a $4\times4$ matrix $\Psi$:
\begin{equation}
    \Psi = \sum_{p=0}^4 \frac{1}{p!}\phi_{\mu_1\dots \mu_p}\gamma^{\mu_1\dots \mu_p},
\end{equation}
where $\gamma^{\mu_1\dots \mu_p}$ is the antisymmetrized product of gamma matrices. Written in terms of the $4\times 4$ KD spinor field $\Psi$, the KD equation (\ref{eq:KD_eq}) becomes 
\begin{equation}\label{eq:KD-Dirac}
    (i\gamma^\mu \partial_\mu - m)\Psi = 0,
\end{equation}
so if we split $\Psi$ into its four columns $\Psi=(\psi_1,\cdots,\psi_4)$, each column obeys the ordinary Dirac equation; and the KD action (\ref{eq:KD_action}) becomes
\begin{equation}\label{eq:KD_action_Psi}
  S_{KD}=\int d^{4}x\,{\rm Tr}[\bar{\Psi}(i\partial\!\!\!\!\;\!/-m)\Psi],\;\;\bar{\Psi}\equiv \gamma^{0}\Psi^{\dagger}\gamma^{0}.
\end{equation}
Note that, compared to the usual definition of $\bar{\Psi}$, this definition has a crucial {\it extra} $\gamma^{0}$ on the {\it left}.

With this extra $\gamma^{0}$, the action is invariant under a Lorentz transformation $x\to \Lambda x$, with $\Psi$ transforming as $\Psi\to U_{\Lambda}^{}\Psi U_{\Lambda}^{-1}$ (where $U_{\Lambda}$ is the $4\times4$ Dirac spinor representation of $\Lambda$).  Since this transformation involves two copies of $U_{\Lambda}$, it has integer spin, and seems bosonic (reflecting its bosonic, $p$-form roots). So is $\Psi$ a boson or a fermion?

To clarify this point, note that the action (\ref{eq:KD_action_Psi}) actually has a larger symmetry ${\rm Spin}(3,1)\times U(2,2)$: {\it i.e.}\ it is invariant under a Lorentz transformation where $\Psi$ transforms as $\Psi\to U_{\Lambda}\Psi V$, where $V$ is any element of the group $U(2,2)=\{V| V\Lambda V^{\dagger}=\Lambda\}$, where $\Lambda={\rm diag}(1,1,-1,-1)$.  So the left-acting $U_{\Lambda}$ corresponds to the Lorentz group (under which $\Psi$ transforms like a {\it spinor}), while the right acting $V$ is a $U(2,2)$ {\it internal} symmetry (since $\Psi\to \Psi V$ is still a symmetry in the absence of a Lorentz transformation). However, when we restrict to the diagonal subgroup $V=U_{\Lambda}^{-1}$ (a certain combination of a Lorentz transformation and an internal symmetry transformation), $\Psi$ transforms {\it as if} it were a polyform field (which is why it lives nicely on a lattice or more general cell complex).  The spinorial nature of $\Psi$ agrees with the fact that, to make sense of the quantization below, we must treat $\Psi$ as an Grassmann-valued field, and quantize using fermionic anti-commutation relations. 

\vspace{0.25cm}
\textbf{Two sheeted spacetime}: The extra $\gamma^{0}$ in the action (\ref{eq:KD_action_Psi}) naively causes problems, since it implies half the fields have ``wrong sign" Lagrangian.  To see this, diagonalize the extra $\gamma^{0}$: $\gamma^{0}=\Omega\Lambda\Omega^{T}$, with $\Lambda={\rm diag}(1,1,-1,-1)$ and $\Omega=(\Omega^{T})^{-1}\in O(4)$.  In terms of $\Psi'\equiv\Psi\Omega$, the action is:
\begin{equation}
  \label{S_KD_Lorentzian1p}
  S_{KD}=\int d^{4}x\,{\rm Tr}[\Lambda\Psi'{}^{\dagger}\gamma^{0}(i\partial\!\!\!\!\;\!/-m)\Psi'].
\end{equation}
In other words, if we split $\Psi'$ into its four columns 
$\Psi'=(\psi_1',\cdots,\psi_4')$, the KD Lagrangian splits into four ordinary Dirac Lagrangians, with overall signs determined by $\Lambda$: the first two (last two) columns have the right (wrong) sign, respectively.

If we follow the standard quantization procedure, such wrong-sign fields lead to negative-norm states (ghosts), so at first sight we might conclude that the theory is simply unphysical. However, as nicely explained in \cite{Donoghue:2019ecz}, an overall minus sign in front of the action corresponds to a reversal of the arrow of causality in quantum theory; or, more correctly, parity and time ($PT$) reversal (see below).  A field with a ``wrong-sign" action should be quantized using anti-time-ordered propagators (rather than time-ordered,  Feynman propagators).

In the present theory, this suggests a picture in which there are two sheets of spacetime -- one with its arrow of causality pointing forward, and the other pointing backward. The $\Lambda_{ii}=+1$ spinors, or the first two columns of $\Psi'$, live on the ``forward" sheet, while the $\Lambda_{ii}=-1$ spinors, or the last two columns in $\Psi'$, live on the ``backward" sheet. Indeed, the two sheets are $PT$ mirror to each other, since an inversion of the coordinates $x^{\mu}\to-x^{\mu}$ on the $\Lambda_{ii}=-1$ sheet flips the Dirac operator $i\partial\!\!\!\!\;\!/\,\to-i\partial\!\!\!\!\;\!/\,$, and hence the action on that sheet, so that it looks identical to the action on the $\Lambda_{ii}=+1$ sheet!

Another (possibly more convenient) way to view it is that the ``two sheets" have the same spacetime orientation, but instead use different square roots of $-1$ (again, see \cite{Donoghue:2019ecz}): where ``our sheet" uses $+i$, the ``other sheet" uses $-i$ (in particular, in the laws of quantum mechanics). For example the plane wave $e^{-ip\cdot x}$ on this sheet corresponds to $e^{ip\cdot x}$ on the other sheet; and quantum mechanics sums over ${\rm e}^{iS}$, canonically quantizes using $[x,p]=i$, or evolves using $\dot{A}=i[H,A]$ on our sheet, but uses ${\rm e}^{-iS}$, $[x,p]=-i$ and $\dot{A}=-i[H,A]$ on the other.

Both pictures fix the KD theory, yielding positive energy, positive norm states on both sheets.

\vspace{0.25cm}
\textbf{KD-Majorana condition}: If ``our" standard model particles live on ``our" sheet, what are the particles on the other sheet? This raises two problems. First, we must avoid any interactions between the two sheets that would violate causality or otherwise conflict with experiment. Second, if we manage to make the two sheets independent yet non-interacting, we would have effectively doubled the number of particles or complexity in the standard model, in a way that could not be tested experimentally.

Fortunately, there is an elegant solution to both problems. 
Whereas the charge conjugate of a Dirac spinor is $\psi^c=i\gamma^{2}\psi^{\ast}$, the charge conjugate of a KD spinor is $\Psi^c=(i\gamma^2)\Psi^{\ast}(i\gamma^2)$ \cite{Catterall:2020fep, Butt:2021brl}; so instead of the usual Majorana condition $\psi_{c}=\psi$, we can impose the ``KD Majorana condition"\footnote{Note that the concept of KD-Majorana condition is similar to the reduced Kähler–Dirac field mentioned in \cite{Butt:2021brl}. In this letter, we further clarify its necessity and physical meaning.}
\begin{equation}
  \Psi^{c}\equiv(i\gamma^{2})\Psi^{\ast}(i\gamma^{2})=\Psi,
\end{equation} 
or, equivalently, in terms of $\Psi'=\Psi\Omega$,
\begin{equation}
  \Psi'{}^{c}\equiv(i\gamma^{2})(\Psi')^{\ast}\beta=\Psi',\quad \beta\equiv \Omega^{T}(i\gamma^{2})\Omega.
\end{equation}
Note that this Majorana condition is Lorentz covariant with respect to both the left-acting ($\Psi\to U_{\Lambda}\Psi$) and left-right-acting ($\Psi\to U_{\Lambda}\Psi U_{\Lambda}^{-1}$) Lorentz transformations.  

With this definition, the KD Majorana condition is a kind of CPT symmetry, since the charge conjugate of a LH (RH) particle in the forward sheet, become RH (LH) anti-particle in the (PT-relatd) backward sheet, and vice versa, in the sense that (for example)
\begin{equation}
    \Psi=\begin{pmatrix} \psi_{1}, \psi_{2},0,0\end{pmatrix}\quad \Psi_c=\begin{pmatrix} 0,0,-\psi_{2}^{c}, \psi_{1}^{c}\end{pmatrix}.
\end{equation}
(Note that $\psi_{1}^c,\psi_2^c$ have swapped positions.) So, if $\Psi'$ satisfies the Kähler–Dirac version of the Majorana condition, its four columns have the form 
\begin{equation}\label{eq:KD-Majorana}
  \Psi'=(\psi_{1}, \psi_{2}, -\psi_{2}^{c}, \psi_{1}^{c}).
\end{equation}

Just as an ordinary Majorana spinor cannot make a particle on its own, but instead creates a LH particle in symmetric combination with its RH anti-particle, a KD-Majorana spinor cannot create a LH particle on the forward sheet on its own, but will always create a LH particle on the forward sheet in combination with a RH anti-particle on the backward sheet, such that their superposition is invariant under $\Psi \to \Psi_c$.  In a sense, we have a sheet and anti-sheet, and the KD Majorana condition requires an invariant combination of the two. 

Note that, from the ordinary (single-sheeted) perspective, if we impose the KD Majorana condition (and regard $\Psi$ as a Grassmann variable), the KD action vanishes identically, because the action from one sheet precisely cancels the action from the other.  But since the two sheets carry two different values of $(-1)$, their contributions to the path integral do not subtract in this way. 

To understand the full symmetry of a single KD Majorana field, write $\Psi'=\begin{pmatrix}a&b\\c&d\end{pmatrix}$ in $2\times2$ block form,  
so the KD action (\ref{S_KD_Lorentzian1p}) (with $m=0$) becomes
\begin{equation}
  S_{KD}=\int d^{4}x{\rm Tr}[a^{\dagger}\bar{\partial}\!\!\!\!\;\!/\;\!a-b^{\dagger}\bar{\partial}\!\!\!\!\;\!/\;\!b+c^{\dagger}\partial\!\!\!\!\;\!/\;\!c-d^{\dagger}\partial\!\!\!\!\;\!/\;\!d]
\end{equation}
with $\partial\!\!\!\!\;\!/=\sigma^{\mu}\partial_{\mu}$, 
$\bar{\partial}\!\!\!\!\;\!/=\bar{\sigma}^{\mu}\partial_{\mu}$.
First note that, before imposing the KD Majorana condition, this actually has a larger internal symmetry than we described previously: for $(U_{\Lambda},V_{L},V_{R})\in Spin(3,1)\times U(2,2)_{L}\times U(2,2)_{R}$, it is symmetric under
\begin{eqnarray}
  (a\;\;b)&\to& U_{\Lambda}^{L}(a\;\;b)V_{L}, \nonumber\\
  (c\;\;b)&\to& U_{\Lambda}^{R}(c\;\;d)V_{R}
\end{eqnarray}
where $V_{L}$ and $V_{R}$ are independent $4\times 4$ matrices, and $U_{\Lambda}^{L}$ and $U_{\Lambda}^{R}$ are the upper and lower $2\times2$ irreducible blocks of $U_{\Lambda}$.  But after imposing the KD Majorana condition (which forces $d=-\sigma^{2}a^{\ast}\sigma^{2}$ and $c=\sigma^{2}b^{\ast}\sigma^{2}$), we must impose $U_{\Lambda}^{R}=(i\sigma^{2})U_{\Lambda}^{L\ast}(i\sigma^{2})$ (which was already true) and $V_{R}=(i\gamma^{2})V_{L}^{\ast}(i\gamma^{2})$, so that the two internal $U(2,2)$'s are no longer independent, and the symmetry is reduced to $Spin(3,1)_{{\rm Lorentz}}\times U(2,2)_{{\rm internal}}$.

\vspace{0.25cm}
\textbf{Relation to the Standard Model (SM)}: On the one hand, a single SM generation (including a RH neutrino) is described by 16 Weyl spinor fields.  On the other hand, a single KD field satisfying the KD-Majorana condition contains two Dirac spinors, or four Weyl spinors. As a result, four KD Majorana fields are required to represent a full generation. We borrow the idea from the Pati-Salam  unification scheme \cite{Pati:1974yy} and its implementation via KD fermions \cite{Catterall:2023nww} to allocate leptons and quarks into the four copies $\Psi'_{\!A=0,\ldots,3}$.

The first copy represents leptons: 
\begin{equation}\label{eq:Psi_leptons}
  \Psi'_{0}=\chi_{0}+\chi_{0}^{c},\text{where }  \chi_{0}=\left(\begin{array}{cccc}
    \nu_{L} & e_{L} & 0 & 0 \\
    \nu_{R} & e_{R} & 0 & 0 \end{array}\right),
\end{equation} 
where all the entries in this matrix are 2-component Weyl spinors.  The next three copies ($\Psi'_{i=1,2,3}$) represent the three colors of quarks:
 \begin{equation}\label{eq:Psi_quarks}
   \Psi'_{i}=\chi_{i}+\chi_{i}^{c},\text{where }\chi_{i}=\left(\begin{array}{cccc}
    u_{L}^{i} & d_{L}^{i} & 0 & 0 \\
    u_{R}^{i} & d_{R}^{i} & 0 & 0 \end{array}\right),
 \end{equation}

Let us see how we can rewrite all the fermion terms in the standard model in this language.

The SM fermion kinetic terms neatly unify as
\begin{equation}
    \mathcal{L}_{\text{KD}} = \text{Tr}[\bar{\Psi}_{\!A}(i\partial\!\!\!\!\;\!/\,)\Psi_{\!A}]=
    \text{Tr}[\Lambda \Psi_{\!A}'^{\dagger}\gamma^{0}(i\partial\!\!\!\!\;\!/\,)\Psi_{\!A}'],
\end{equation}
where $A=0,\ldots,3$ is summed over.  In addition to the $Spin(3,1)_{{\rm Lorentz}}\times U(2,2)_{{\rm internal}}$ symmetry described in the previous section, this action now also has an internal $SU(4)$ symmetry permuting the four copies of $\Psi'$ (in the fundamental irrep of $SU(4)$).  Gauging an $SU(3)$ subgroup of $SU(4)$, and an $SU(2)\times U(1)$ subgroup of $U(2,2)$ (replacing $\partial\!\!\!\!\;\!/$ by the appropriate covariant derivative) yields the standard model fermion kinetic terms (including gauge interactions).

\vspace{0.25cm}
\textbf{Yukawa couplings}: Having outlined the kinetic structure, we now rewrite the Yukawa couplings in the KD formulation. 
Recall that the standard Yukawa Lagrangian ${\cal L}_{Y}$ (including a right-handed neutrino) is as follows:
\begin{equation}\label{eq:Yukawa_standard}
     \tilde{h}^{T}\!\bar{q}_L^{\,i}u_R^j y_u^{ij} + 
     h^{T}\!\bar{q}_L^{\,i}d_R^j y_d^{ij} + \tilde{h}^{T\;\!}\!\bar{\ell}_L^{\,i}\nu_R^j y_\nu^{ij} + h^{T\;\!}\!\bar{\ell}_L^{\,i} e_R^j y_e^{ij} + \text{h.c.}
\end{equation}
where $h$ is the Higgs doublet and $\tilde{h}\equiv i\sigma^{2}h^{\ast}$; 
$y_{u}^{ij}$, $u_{d}^{ij}$, $y_{\nu}^{ij}$ and $y_{e}^{ij}$ are all complex $3\times 3$ Yukawa matrices (where $i,j=1,2,3$ are the generation indices);
$q_{L}=(u_{L},d_{L})$ and $\ell_{L}=(\nu_{L},e_{L})$ are the left-handed quark and lepton doublets (written as $2\times2$ matrices); and, in this section, we revert to the standard (non-KD) bar notation for spinors ({\it e.g.}\ $\bar{q}_{L}=q_{L}^{\dagger}\gamma^{0}$, without a extra $\gamma^{0}$ on the left).

The Yukawa couplings may be neatly compressed in the KD formalism as follows:
\begin{equation}\label{eq:Yukawa_KD}
    \mathcal{L}_{Y}=
    \text{Tr}[\mathcal{H}^{T}\widehat{\Psi}_{A,L}'^{i}\Psi_{A,R}'^j \mathcal{Y}_A^{ij}] + \text{h.c.}  
\end{equation}
where $\widehat{\Psi}' \equiv \Lambda \Psi'^\dagger \gamma^0$ and 
the KD Higgs field ${\cal H}$ and Yukawa couplings ${\cal Y}^{ij}_{A}$ are explained below.

Let us demonstrate how the above expression works explicitly for the lepton sector ($A=0$). The building blocks are the $4 \times 4$ matrices for the lepton field $\Psi_0$, Higgs field $\mathcal{H}$, and Yukawa couplings $\mathcal{Y}_0$:
\begin{gather}
    \Psi_{0,L}'=\begin{pmatrix} \ell_L & 0 \\ 0 & \ell_L^c \end{pmatrix},
    \Psi_{0,R}'=\begin{pmatrix} 0 & \ell_R^c \\ \ell_R & 0 \end{pmatrix}\\
    \mathcal{H} = \begin{pmatrix} H & 0 \\ 0 & H^{c} \end{pmatrix}, \text{where } H=H^c=(\tilde h, h) \\
    \!\!\mathcal{Y}_0^{ij} = \begin{pmatrix}
        Y_0^{ij} & 0 \\ 0 & Y_0^{c,ij}
    \end{pmatrix},
    \begin{aligned}
        Y_0^{ij} &= \text{diag}(y_\nu^{ij}, y_e^{ij}), \\
        Y_0^{c,ij} &= \text{diag}(y_e^{ij*}, y_\nu^{ij*}),
    \end{aligned}
\end{gather}
where $H^c=\sigma^2H^*\sigma^2,Y_0^{c,ij}=\sigma^2Y_0^{ij,*}\sigma^2$.

We then apply them to extend \cref{eq:Yukawa_KD}:
\begin{equation}\label{eq:yukawa_expand}
\begin{aligned}
    &\text{Tr}[\mathcal{H}^{T}\widetilde{\Psi}_{0,L}'^{i}\Psi_{0,R}'^j \mathcal{Y}_0^{ij}] \nonumber \\
    &=\text{Tr}[H^{T}\bar{\ell}_{L}^{\,i}\ell_{R}^{\,j} Y_0^{ij}] - \text{Tr}[H_{c}^{T}\bar{\ell}_{L}^{\,c,i}\ell_{R}^{\,c,j} Y_0^{c,ij}].
\end{aligned}
\end{equation}
The first/second term represents the Yukawa couplings on the forward/backward sheet.

To see this, let us further expand the $2\times 2$ components. The first term $\text{Tr}[H^{T}\bar{\ell}_{L}^{\,i}\ell_{R}^jY_0^{ij}]$ becomes: 
\begin{equation}
\begin{aligned}
    \text{Tr}[H^{T}\bar{\ell}_{L}^{\,i}\ell_{R}^jY_0^{ij}]
    =\tilde{h}^{T}\bar{\ell}_{L}^{\,i} \nu_R^j y_\nu^{ij}+h^{T} \bar{\ell}_{L}^{\,i}e_R^j y_e^{ij}.
\end{aligned}
\end{equation}
This reproduces the two Yukawa coupling terms for the lepton sector. Similarly, by expanding the second term in \cref{eq:yukawa_expand}, we get the corresponding terms for anti-particles on the ``other" sheet. 

Following the same method, the calculation for the quark sector ($\Psi_i$) generates all remaining Yukawa couplings of the Standard Model. Thus, (\ref{eq:Yukawa_KD}) successfully unifies all Yukawa interactions.

\vspace{0.25cm}
\textbf{Neutrino Majorana mass term}: The last fermion term in the standard model (when we include RH neutrinos) is the neutrino Majorana mass term ${\cal L}_{M}$, which takes the form
\begin{equation}
  {\cal L}_{M}=\bar{\nu}_R^{\,c,i}\nu_{R}^{\,j}m^{ij}+\text{h.c.}
\end{equation}
where $m^{ij}$ is a symmetric $3\times3$ matrix that, without loss of generality, may be made real by appropriate field redefinitions.  In the KD formalism, this takes the form

\begin{equation}
  {\cal L}_{M}={\rm Tr}[\widehat{\Psi}_{A,R}'^{i}\Psi_{A,R}'^{j}\mathcal{M}_{A}^{ij}]
\end{equation}
where gauge invariance restricts the Majorana mass matrix ${\cal M}_{A}^{ij}$ to the form
\begin{equation}
  {\cal M}_{A}^{ij}= \begin{pmatrix}0 & M_{c}^{ij} \\
  M^{ij} & 0 \end{pmatrix}
  \delta_{A}^{0},\quad
  M^{ij}=\begin{pmatrix} 0 & 0 \\
  m^{ij} & 0 \end{pmatrix}.
\end{equation}

Note that, just as for the free KD theory above, the KD Majorana condition implies that in all the standard model fermion terms (kinetic, Yukawa, and Majorana mass), the actions on the two sheets are precisely equal ({\it i.e.}\ they naively cancel from a one-sheeted perspective). 

\vspace{0.25cm}
\textbf{Discussion and Future work}: It is natural to wonder about the physical implications of this two sheeted spacetime picture (with a KD-Majorana symmetry relating the two sheets).

In particular, it is intriguing to speculate about implications for cosmology. One can imagine the forward and backward sheets are ``after" and ``before" the Big Bang, and are joined at bang itself \cite{Boyle:2021jej}. This picture seems strikingly related to the CPT-Symmetric Universe picture explained in \cite{Boyle:2018tzc, Boyle:2018rgh, Boyle:2021jej, Turok:2022fgq, Boyle:2022lcq}, in which the Big Bang is a mirror at which two sheets of spacetime (a universe/anti-universe pair, related by CPT symmetry) are joined.  This cosmological picture was originally motivated by the observed properties of the primordial universe (in particular, the reflecting boundary conditions that cosmological perturbations are observed to obey when we follow them back to the beginning of the radiation era \cite{Boyle:2018tzc}). But {\it why} did they obey reflecting boundary conditions at early times?  That is {\it why} was the Big Bang a CPT mirror?  It seems the KD Majorana condition suggested here (in order to make sense of KD fermions) provides an elegant explanation.

Relatedly, the ``black mirror" was recently proposed alternative to the standard black hole. In the new picture, the black hole interior does not exist, and the black hole horizon is instead a mirror connecting our exterior to the ``other" exterior ({\it i.e.}\ another mirror connecting our universe to its anti-universe) \cite{tzanavaris2025blackmirrorscptsymmetricalternatives}, so each infalling particle from our universe annihilates with its corresponding anti-particle from the anti-universe at the horizon. The KD Majorana condition explains why infalling particles are paired up in this way.

A forthcoming companion paper \cite{BoyleVaibhav} will explain how KD fermions also suggest a new approach (building on \cite{Catterall:2022jky, Catterall:2023nww}) to putting chiral gauge theory on the lattice.

{\bf Note}: Shortly before posting this paper, we became aware that Catterall and Hubisz are also preparing a forthcoming related paper (on quantizing the Lorentzian Kahler-Dirac theory). Their picture is different but has some overlap with ours.

\vspace{0.25cm}
\textbf{Acknowledgements}:
We are grateful to Simon Catterall, Jay, Hubisz, Sotiris Mygdalas, Neil Turok, and Vatsalya Vaibhav for valuable and insightful comments and discussion, and to Catterall and Hubisz for sharing their unpublished notes.  LB is supported by the STFC Consolidated Grant ‘Particle Physics at the Higgs Centre’. WND is supported by a Cambridge Trusts Taiwanese Scholarship. Research at the Perimeter Institute is supported by the Government of Canada, through Innovation, Science and Economic Development, Canada and by the Province of Ontario through the Ministry of Research, Innovation and Science.  

\bibliography{Reference}

\end{document}